\documentclass[fleqn,usenatbib]{mnras}
\pdfoutput=1
\usepackage{graphicx}
\usepackage{amsmath,amssymb,amstext}
\usepackage{lastpage}
\usepackage[T1]{fontenc}
\usepackage{txfonts}

\usepackage[all]{hypcap}
\usepackage{subfig}
\newcommand*{\subfigure}{\subfloat}

\newcommand*{\eagle}{EAGLE}
\newcommand*{\owls}{OWLS}
\newcommand*{\rockstar}{\textsc{Rockstar}}
\newcommand*{\subfind}{\textsc{Subfind}}

\title{On the galaxy--halo connection in the EAGLE simulation}
\author[H.~Desmond, Y.-Y.~Mao, R.~H.~Wechsler, R.~A.~Crain and J.~Schaye]{
Harry~Desmond$^{1,2}$\thanks{E-mail: harryd2@stanford.edu},
Yao-Yuan~Mao$^{3,4}$,
Risa~H.~Wechsler$^{1,2}$,
Robert~A.~Crain$^5$,
and \newauthor{}
Joop~Schaye$^{6}$
\\
$^{1}$Kavli Institute for Particle Astrophysics and Cosmology, Physics Department, Stanford University, Stanford, CA 94305, USA\\
$^{2}$SLAC National Accelerator Laboratory, Menlo Park, CA 94025, USA\\
$^{3}$Department of Physics and Astronomy, University of Pittsburgh, Pittsburgh, PA 15260, USA\\
$^{4}$Pittsburgh Particle Physics, Astrophysics, and Cosmology Center (PITT PACC), Pittsburgh, PA 15260, USA\\
$^{5}$Astrophysics Research Institute, Liverpool John Moores University, 146 Brownlow Hill, Liverpool, L3 5RF, UK\\
$^{6}$Leiden Observatory, Leiden University, P.O. Box 9513, 2300 RA Leiden, the Netherlands
}

\pubyear{2017}

\begin{document}
\label{FirstPage}
\pagerange{\pageref{FirstPage}--\pageref{LastPage}}
\maketitle

\begin{abstract}
Empirical models of galaxy formation require assumptions about the correlations between galaxy and halo properties. These may be calibrated against observations or inferred from physical models such as hydrodynamical simulations. In this \textit{Letter}, we use the \eagle{} simulation to investigate the correlation of galaxy size with halo properties. We motivate this analysis by noting that the common assumption of angular momentum partition between baryons and dark matter in rotationally supported galaxies overpredicts both the spread in the stellar mass--size relation and the anticorrelation of size and velocity residuals, indicating a problem with the galaxy--halo connection it implies. We find the \eagle{} galaxy population to perform significantly better on both statistics, and trace this success to the weakness of the correlations of galaxy size with halo mass, concentration and spin at fixed stellar mass. Using these correlations in empirical models will enable fine-grained aspects of galaxy scalings to be matched.
\end{abstract}


\section{Introduction}
\label{sec:intro}

Accurate semi-analytic and empirical modelling of galaxy formation is challenging, in part because the correlations of key galaxy and halo variables remain unknown. Observational manifestations of these correlations include galaxy scaling relations, and through detailed investigations of these relations we may hope to build knowledge of the galaxy--halo connection.

Over the past decades, two models which have proven useful for capturing aspects of the galaxy--halo connection are subhalo abundance matching (SHAM; \citealt{Kravtsov, Behroozi}), and the angular momentum model of \citet*[hereafter MMW]{MMW}. SHAM asserts a nearly monotonic relationship between stellar mass and a halo proxy, establishing the dependence of galaxy mass on halo mass and concentration required to reproduce galaxy clustering~\citep*[e.g.][]{Conroy, Reddick}. The MMW model sets galaxy and halo specific angular momentum proportional and assumes galaxies' velocities to be entirely rotational, making galaxy size a function of galaxy mass and halo mass, concentration and spin. This agrees well with observed average galaxy sizes over a wide range of mass (\citealt{Kravtsov_Radius, DW15}, hereafter DW15).

Despite these successes, however, the conjunction of these models (hereafter ``SHAM+MMW'') is known to make incorrect predictions for two properties of the galaxy population. The first is the scatter $s_\text{MSR}$ in the stellar mass--size relation (MSR;~\citealt{deJong, Gnedin_new}). SHAM+MMW sets galaxy size proportional to halo spin, $\lambda$, and hence requires the scatter in size at fixed mass to be at least as large as that in $\lambda$. In fact, these scatters are $\sim0.2$ dex and $\sim0.25$ dex in observed galaxies and simulations respectively (DW15). The second is the correlation of residuals of the mass--size and mass--velocity relations ($\rho_{\Delta R-\Delta V}$), which is negligible in observations but predicted to be negative (\citealt{McGaugh_Residuals, D07}; DW15). These discrepancies indicate that the galaxy--halo correlations on which $s_\text{MSR}$ and $\rho_{\Delta R-\Delta V}$ depend are inadequately captured by the model.

This issue is relevant also for semi-analytic models. Many such models set galaxy size proportional to halo virial radius (e.g.~\citealt{Somerville, Croton, Lu11}), sometimes with a single value for all halo spins. Others that use additional physical assumptions find important correlations of size with variables beyond halo mass and spin, but neglect the scatter in sizes (e.g.~\citealt*{Lu15}). The empirical identification of the aspects of the galaxy--halo connection responsible for realistic size distributions -- and correlations with velocity -- will be of use in constraining such models and guiding the choice of inputs.

The failure of SHAM+MMW may be due either to inaccuracies in the properties of the halo populations on which the models were based (e.g. their neglect of baryonic physics), or incorrect prediction of the models themselves for the galaxy--halo connection. To resolve this dilemma, we turn in this \textit{Letter} to hydrodynamical simulations, which enable the prediction of galaxy properties without prior assumptions on galaxy--halo correlations. In particular, we investigate $s_\text{MSR}$ and $\rho_{\Delta R-\Delta V}$ in the \eagle{} simulation~\citep{EAGLE, EAGLE_2},\footnote{\url{http://eagle.strw.leidenuniv.nl}} which has previously been shown to match the galaxy size distribution as well as many other aspects of galaxy phenomenology~\citep{EAGLE_sizes}. \citet{Sales09} and~\citet{EAGLE_MMW} showed that the MMW model fails to match the output of the \eagle{} simulation and its ancestor \owls{}. \citet{Zavala} found the angular momentum of stars to correlate with that of the inner halo in \eagle{}, and~\citet{Sales} reported weak correlation of galaxy properties with halo spin in the related GIMIC simulation. This is in contrast with other simulations in which halo spin correlates more strongly with galaxy spin and morphology, especially at low mass (e.g.~\citealt{Teklu,Rodriguez-gomez}). Finally,~\citet{Ferrero} studied the \eagle{} Tully--Fisher and mass--size relations.

The structure of this paper is as follows. In Section~\ref{sec:method} we describe the \eagle{} simulation and our methods to measure and explore the origin of $s_\text{MSR}$ and $\rho_{\Delta R-\Delta V}$. In Section~\ref{sec:observables} we show that both $s_\text{MSR}$ and $\rho_{\Delta R-\Delta V}$ are significantly nearer their observed values in \eagle{} than in the SHAM+MMW model, and close to the predictions of SHAM alone. We show the success of \eagle{} over SHAM+MMW to be due not to differences in underlying halo properties caused by baryons (Section~\ref{sec:halo_comp}), but rather to the correlations of halo variables with galaxy size (Section~\ref{sec:galhal_comp}). In \eagle{}, the sizes of low-redshift galaxies are only weakly correlated at fixed stellar mass with the mass, concentration and spin of their haloes, violating the assumption of angular momentum partition. Section~\ref{sec:conclusion} discusses the broader implications of these results, and summarises.

\section{Simulations and Methods}
\label{sec:method}

\subsection{The \eagle{} simulation}
\label{sec:eagle}

\eagle{} is a recently completed set of cosmological hydrodynamical simulations, run with a modified version of \textsc{Gadget-3}~\citep{Springel} and including hydrodynamics, radiative cooling, star formation, stellar feedback and black hole dynamics. The subgrid models were calibrated against the present-day stellar mass function and the normalisation of the mass--size relation. The simulations used a flat $\Lambda$CDM cosmology with $\Omega_\text{m} = 0.307$, $\Omega_\text{b} = 0.04825$, $h=0.6777$, $\sigma_8 = 0.8288$ and $n_\text{s} = 0.9611$. We analyse the $z=0$ snapshot of simulation Ref-L100N1504, which tracks $1504^3$ dark matter and gas particles from $z=127$ to the present day in a box with comoving side length 100~Mpc, in addition to the corresponding dark matter only (DMO) run in which baryonic effects were switched off. We refer the reader to~\citet{EAGLE} and~\citet{EAGLE_2} for further information about the simulation.

\subsection{Finding and matching haloes}
\label{sec:matching}

To enable direct comparison with the results of DW15, we perform halo finding on both the DMO and hydrodynamical (hereafter ``hydro'') runs of the \eagle{} simulation using \rockstar{}~\citep{rockstar}. We define spin as $\lambda \equiv J|E|^{1/2}G^{-1}M^{-5/2}$, where $J$ is a halo's angular momentum and $E$ its total energy~\citep{Peebles1969}, and calculate concentration ($c$) using $r_{s,\,\text{klypin}}$ (derived from $V_\text{max}/V_\text{vir}$;~\citealt{Klypin2001}) rather than fitting an NFW profile. We include only dark matter when calculating $c$ and $\lambda$. We multiply the DMO haloes' virial masses by $1-\Omega_\text{b}/\Omega_\text{m}$ to compare to the hydro haloes, where again we include dark matter only ($M_\text{DM}$).

Next, we match the DMO \rockstar{} catalogue to both the hydro \rockstar{} catalogue and the \subfind{} catalogue~\citep{Subfind_1,Subfind_2} made by the \eagle{} pipeline, as follows. Both halo finders produce a list of particles associated with each halo that they identify. Since the two runs share the same dark matter particle IDs, we can match the haloes by finding common particles. In practice, given a halo in the DMO run (halo A), we first find the halo (halo B) in the hydro run that contains the most particles of halo A. If halo A also contains the most particles of halo B, we identify a ``match'' between them. Since the \subfind{} catalogue of the hydro run also provides the connection between the haloes and galaxies, this method establishes a link between the haloes in the DMO and hydro \rockstar{} catalogues, and the galaxies in the \subfind{} catalogue. The fraction of haloes in the hydro run hosting galaxies with $M_* > 10^9 \mathrm{M}_\odot$ that are matched by our procedure is $91$ per cent; these haloes are not significantly biased in $M_\text{DM}$, $c$ or $\lambda$.

\subsection{Data, models and statistics}
\label{sec:statistics}

We compare our models with the observations of~\citet[hereafter P07]{P07} for compatibility with DW15. Although larger samples with well-measured sizes now exist (e.g.~\citealt{Huang},~\citealt{Somerville_MSR}), they produce similar MSRs. P07 require an apparent axis ratio $b/a \leq 0.6$ and usable H$\alpha$ rotation curve, which they find not to significantly bias the admitted galaxy population in colour or concentration. We therefore do not make a morphology cut on the \eagle{} galaxies in our fiducial analysis, although we have checked that our results change at no more than the $\sim 1 \sigma$ level -- and our qualitative conclusions remain unchanged -- when only including galaxies with a substantial fraction of their kinetic energy in ordered corotation ($\kappa_\text{co} \geq 0.4$;~\citealt{Correa}). We compare the \eagle{} results with two semi-empirical models, denoted ``SHAM'' and ``SHAM+MMW'' as in Section~\ref{sec:intro}.

For both data and models, we take $s_\text{MSR}$ to be the Gaussian scatter in radius of the best-fitting power-law relation\footnote{Note that the use of a power law in this definition means that $s_\text{MSR}$ is increased by curvature in the MSR; thus $s_\text{MSR}$ for the \eagle{} relation (see Fig.~\ref{fig:MSR}) may be considered an upper bound on the ``true'' intrinsic scatter.} between stellar mass ($M_*$) and half-mass radius ($R_\text{eff}$; measured for stars in a 30~kpc aperture), over the range $9 < \log(M_*/\text{M}_\odot) < 11.5$. We have verified that restricting to $\log(M_*/\text{M}_\odot) > 10$ does not affect our conclusions. We measure $\rho_{\Delta R-\Delta V}$ as the Spearman rank correlation coefficient of the $\Delta R_\text{eff}-\Delta V_\text{max}$ relation, where $\Delta x$ denotes the residual of quantity $\log(x)$ after subtracting the value expected at that $M_*$ given a power-law fit to the $\log(M_*)-\log(x)$ relation, $f_x(M_*)$:

\begin{equation}
\Delta x \equiv \log(x) - f_x(M_*).
\end{equation}

\noindent We quantify the dependence of $R_\text{eff}$ on halo variables with the Spearman correlation coefficients $\rho_{\Delta R-\Delta X}$, where $X \in \{M_\text{DM}, c, \lambda\}$. We record in Table~\ref{tab:table} the median and $1\sigma$ spread of the statistics over 100 Monte Carlo mock data sets of galaxies with $M_*$ values within 0.01 dex of those of the observational sample (see Section~\ref{sec:observables}).

\section{Results}
\label{sec:results}

\subsection{The \eagle{} mass--size and $\Delta{R}-\Delta{V}$ relations}
\label{sec:observables}

\begin{figure*}
  \subfigure[$M_*-R_\text{eff}$]
  {
    \includegraphics[width=0.42\textwidth]{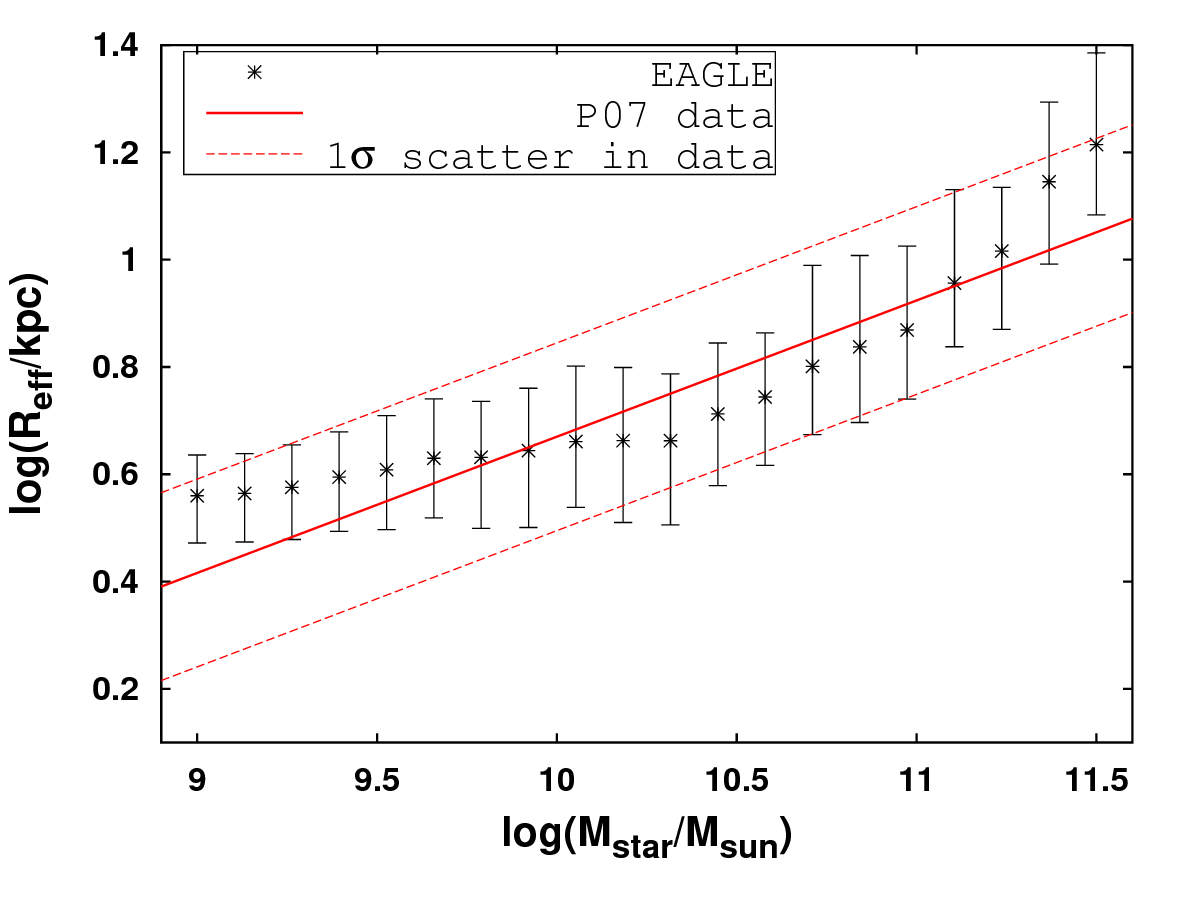}
    \label{fig:MSR}
  }
  \subfigure[$\Delta R_\text{eff} - \Delta V_\text{max}$]
  {
    \includegraphics[width=0.42\textwidth]{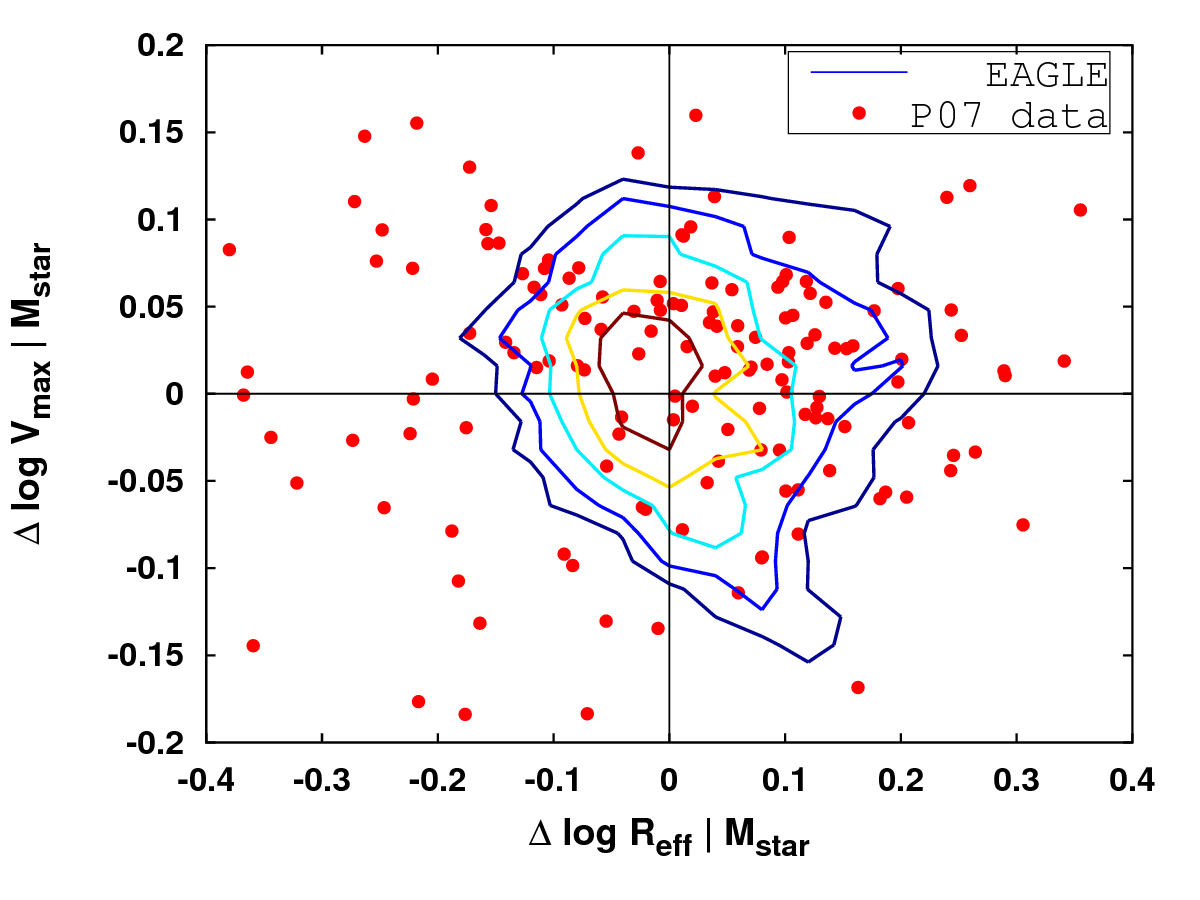}
    \label{fig:dR-dV}
  }
  \caption{The $M_*-R_\text{eff}$ relation and correlation of $R_\text{eff}$ and $V_\text{max}$ residuals in the \eagle{} simulation, compared to the observations of~\citet{P07}. As in DW15, stellar masses for the latter were taken from the NASA Sloan Atlas. The red lines in the left panel show the best-fitting power-law to the data, and its scatter. In this plot and those that follow, points indicate medians and error bars $16^\text{th}$ and $84^\text{th}$ percentiles. We stack the 100 mock data sets (Section~\ref{sec:statistics}) to make contour plots, and the levels enclose 90, 80, 60, 40 and 20 per cent of galaxies. The spread in the sizes of \eagle{} galaxies is as low as is observed, and they correctly exhibit no significant $\Delta R_\text{eff}-\Delta V_\text{max}$ correlation. In Fig.~\ref{fig:dR-dV}, the observations have $\rho_{\Delta R-\Delta V}=-0.07$, and the \eagle{} galaxies have $\rho_{\Delta R-\Delta V}=-0.13$.}
  \label{fig:observables}
\end{figure*}

Figure~\ref{fig:MSR} shows the MSR of the \eagle{} galaxies, and Figure~\ref{fig:dR-dV} the correlation of their size and velocity residuals. That both \eagle{} relations are in approximate agreement with the P07 observations is verified quantitatively in the first two rows of Table~\ref{tab:table}, which list the $s_\text{MSR}$ and $\rho_{\Delta R-\Delta V}$ values.

\begin{table}
  \begin{center}
    \begin{tabular}{l|r|r|r|r}
      \hline
							& P07	        &\eagle{}		&SHAM			&SHAM+MMW\\ 
      \hline
\rule{0pt}{3ex}
      $s_\text{MSR}$ 					& 0.18		& $0.15 \pm 0.01$	& \textit{0.18}		& $0.39 \pm 0.03$\\
\rule{0pt}{3ex}
      $\rho_{\Delta R-\Delta V}$			& $-0.07$	& $-0.13 \pm 0.07$	& $-0.22 \pm 0.04$	& $-0.56 \pm 0.05$\\
\rule{0pt}{3ex}
      $\rho_{\Delta R-\Delta M_\text{DM}}$		& -- 		& $0.18 \pm 0.07$	& \textit{0}		& $0.75 \pm 0.04$\\
\rule{0pt}{3ex}
      $\rho_{\Delta R-\Delta c}$			& -- 		& $-0.19 \pm 0.07$	& \textit{0}		& $-0.76 \pm 0.04$\\
\rule{0pt}{3ex}
      $\rho_{\Delta R-\Delta \lambda}$	         	& -- 		& $0.17 \pm 0.08$       & \textit{0}		& $0.80 \pm 0.04$\\
      \hline
    \end{tabular}
  \caption{Comparison of statistics of the galaxy--halo connection in observations (P07), the \eagle{} simulation, an abundance matching model with sizes chosen to match the stellar mass--size relation by construction (``SHAM''), and an analogous model with sizes set by angular momentum partition (``SHAM+MMW''; \citealt{DW15}). $s_\text{MSR}$ is the scatter in size of the stellar mass--size relation, $\rho$ denotes Spearman rank correlation coefficient, and $\Delta$ is defined in Eq. 1. Entries in italics are by construction. The SHAM+MMW model overpredicts both $s_\text{MSR}$ and $|\rho_{\Delta R-\Delta V}|$ due to the strong correlations it implies between $R_\text{eff}$ and $M_\text{DM}$, $c$ and $\lambda$ at fixed $M_*$. The \eagle{} galaxy--halo connection, in which these variables are only weakly correlated, performs significantly better on both statistics.}
  \label{tab:table}
  \end{center}
\end{table}

The $3^\text{rd}$ and $4^\text{th}$ columns of Table~\ref{tab:table} show analogous results for two alternative models. In ``SHAM,'' $M_*$ is set by SHAM using the $V_\text{peak}$ proxy and 0.2 dex scatter (\citealt{Reddick}; varying the SHAM parameters within the bounds set by clustering has a negligible effect on our results), and galaxy sizes are chosen randomly from a normal distribution at given $M_*$ to match the P07 MSR by construction. In ``SHAM+MMW,'' sizes are set by the MMW model after SHAM has been performed, using the procedure and best-fitting parameter values of DW15.

As mentioned in Section~\ref{sec:intro} (and discussed in detail in DW15), the SHAM+MMW model compares poorly with observations in both $s_\text{MSR}$ and $\rho_{\Delta R-\Delta V}$. This appears to be in conflict with~\citet{Somerville_MSR}, who claim the model generates an $s_\text{MSR}$ in agreement with that of a compilation of GAMA and CANDELS data. However, they include only the contribution to $s_\text{MSR}$ from scatter in $\lambda$ ($\sim0.25$ dex) and neglect the contributions from scatter in $M_\text{DM}$ and $c$ at fixed $M_*$. $\rho_{\Delta R-\Delta V}$ has contributions both from baryonic mass (higher surface density means larger rotation velocity), and from the dark matter, since in the MMW model more concentrated haloes, which generate larger rotation velocities, host smaller galaxies at fixed angular momentum. The SHAM model, which includes only the first contribution, predicts a $\Delta R-\Delta V$ anticorrelation that is weaker but still stronger than the data's. It is important to note, however, that these models assume negligible velocity dispersion $\sigma$. An decrease of $\sigma/V_\text{rot}$ with $\lambda$ -- as produced in some hydrodynamical simulations (e.g.~\citealt{Rodriguez-gomez}) -- could reduce the predicted $|\rho_{\Delta R-\Delta V}|$ and $s_\text{MSR}$. Only with the assumption that $\sigma/V_\text{rot}$ does not vary systematically with $\lambda$ does the MMW model follow uniquely from proportionality of galaxy and halo specific angular momentum.

We now investigate the origin of the difference between the \eagle{} and SHAM+MMW results.

\subsection{Comparison of the haloes in the DMO and hydrodynamical runs of the EAGLE simulation}
\label{sec:halo_comp}

\begin{figure*}
  \subfigure[$M_\text{DM}$]
  {
    \includegraphics[width=0.34\textwidth]{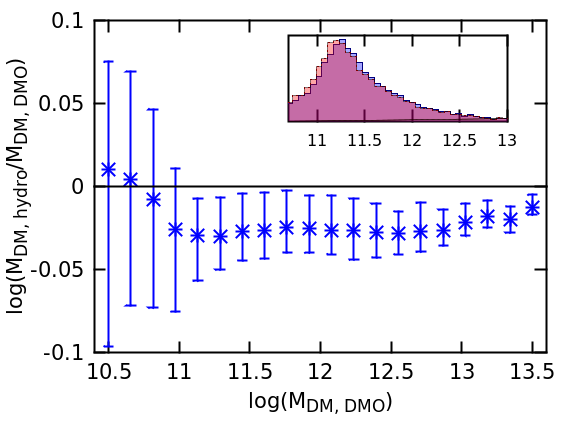}
    \label{fig:rs_mvir}
  }
  \subfigure[$c$]
  {
    \includegraphics[width=0.34\textwidth]{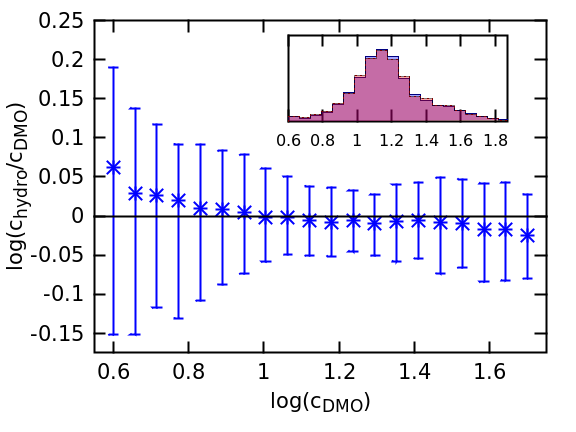}
    \label{fig:rs_conc}
  }
  \subfigure[$\lambda$]
  {
    \includegraphics[width=0.34\textwidth]{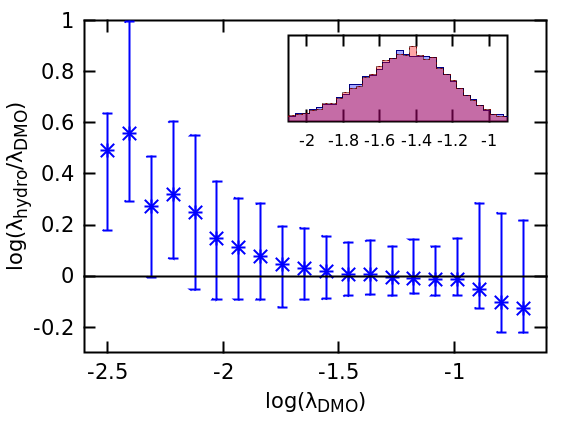}
    \label{fig:rs_spin}
  }
  \caption{The differences in $M_\text{DM}$, $c$ and $\lambda$ between all $M_* > 10^9 \: M_\odot$ haloes in the hydro runs of the \eagle{} simulation and their counterparts in the DMO run, as a function of the DMO variable. The insets compare the overall distributions (hydro in red and DMO in blue). With baryonic effects included, $M_\text{DM}$ is reduced by a few per cent on average (the catalogue is incomplete for $M_\text{DM} \lesssim 10^{11}\:M_\odot$), and $\lambda$ increased slightly at low values. $c$ is largely unaffected.}
  \label{fig:rs}
\end{figure*}

A possible reason for the apparent failure of the SHAM+MMW model is its application in DW15 to haloes from an N-body simulation in which baryonic effects were neglected. In Figure~\ref{fig:rs} we show the fractional differences in $M_\text{DM}$, $c$ and $\lambda$ of all matched haloes in the \eagle{} DMO and hydro runs, and compare in the insets their overall distributions. We find the haloes to be a few per cent less massive on average in the hydro run, and their $c$ and $\lambda$ values to be similar. (\citealt{EAGLE_baryons} reported larger differences in halo mass because they included baryons as well as dark matter in the mass definition.) The spin distribution is slightly wider in the hydro run, which goes in the wrong direction to account for the lower $s_\text{MSR}$ in \eagle{} than in the SHAM+MMW model.

\begin{figure}
  \centering
  \includegraphics[width=0.405\textwidth]{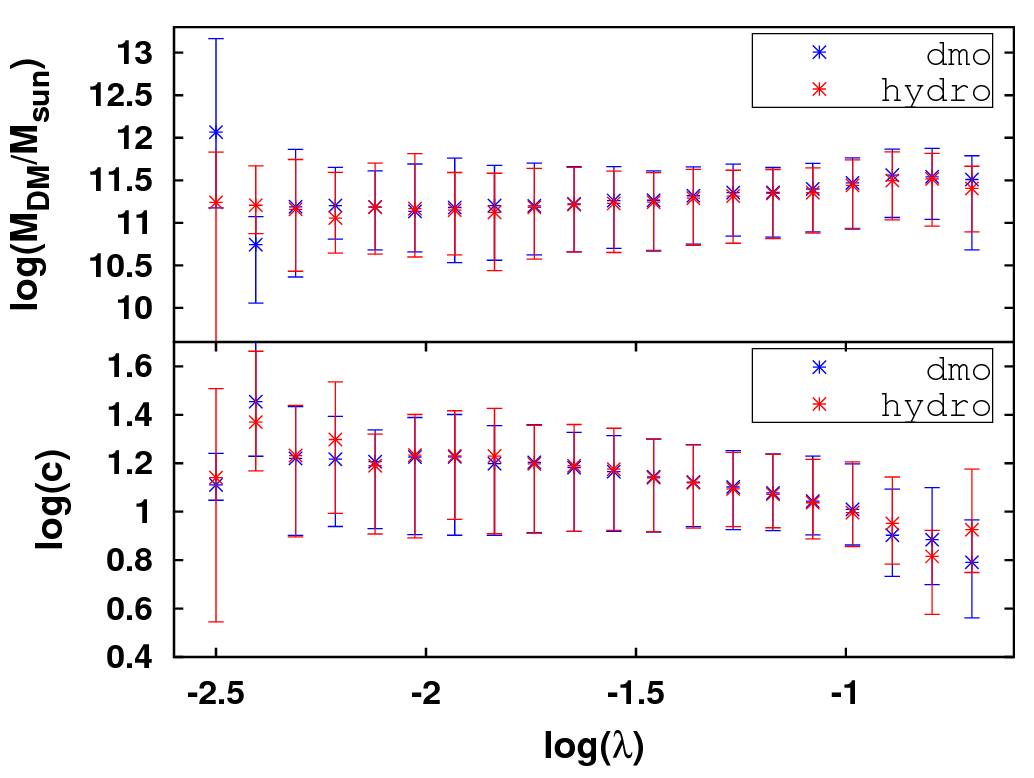}
  \caption{$\lambda$ correlates in the same way with both $M_\text{DM}$ and $c$ in the DMO and hydro runs of \eagle{}. Together with Fig.~\ref{fig:rs}, this shows that the difference between the \eagle{} and SHAM+MMW models in their predictions for $s_\text{MSR}$ and $\rho_{\Delta R-\Delta V}$ are not due to changes to the haloes caused by baryons. They must therefore be due to different galaxy--halo correlations.}
  \label{fig:spin_corr}
\end{figure}

In Figure~\ref{fig:spin_corr} we compare the correlations of $\lambda$ with $M_\text{DM}$ and $c$ in the two runs, finding them to be very similar. If spin was more positively correlated with $M_\text{DM}$ or $c$ with baryonic effects included, then the corresponding increase in rotational velocity caused by dark matter for larger galaxies would compensate for the reduction in the rotation velocity caused by baryons, which could allow the SHAM+MMW model to agree with the measured $\rho_{\Delta R-\Delta V}$. That we do not find such an increased correlation leads us to conclude that the differences between the \eagle{} and SHAM+MMW models in their predictions for $s_\text{MSR}$ and $\rho_{\Delta R-\Delta V}$ arise not from underlying dark matter halo structure, but rather from differences in the correlations of galaxy and halo variables. It is to these that we now turn.

\subsection{The galaxy--halo connection}
\label{sec:galhal_comp}

\begin{figure*}
  \subfigure[$\Delta R_\text{eff}-\Delta M_\text{DM}$]
  {
    \includegraphics[width=0.33\textwidth]{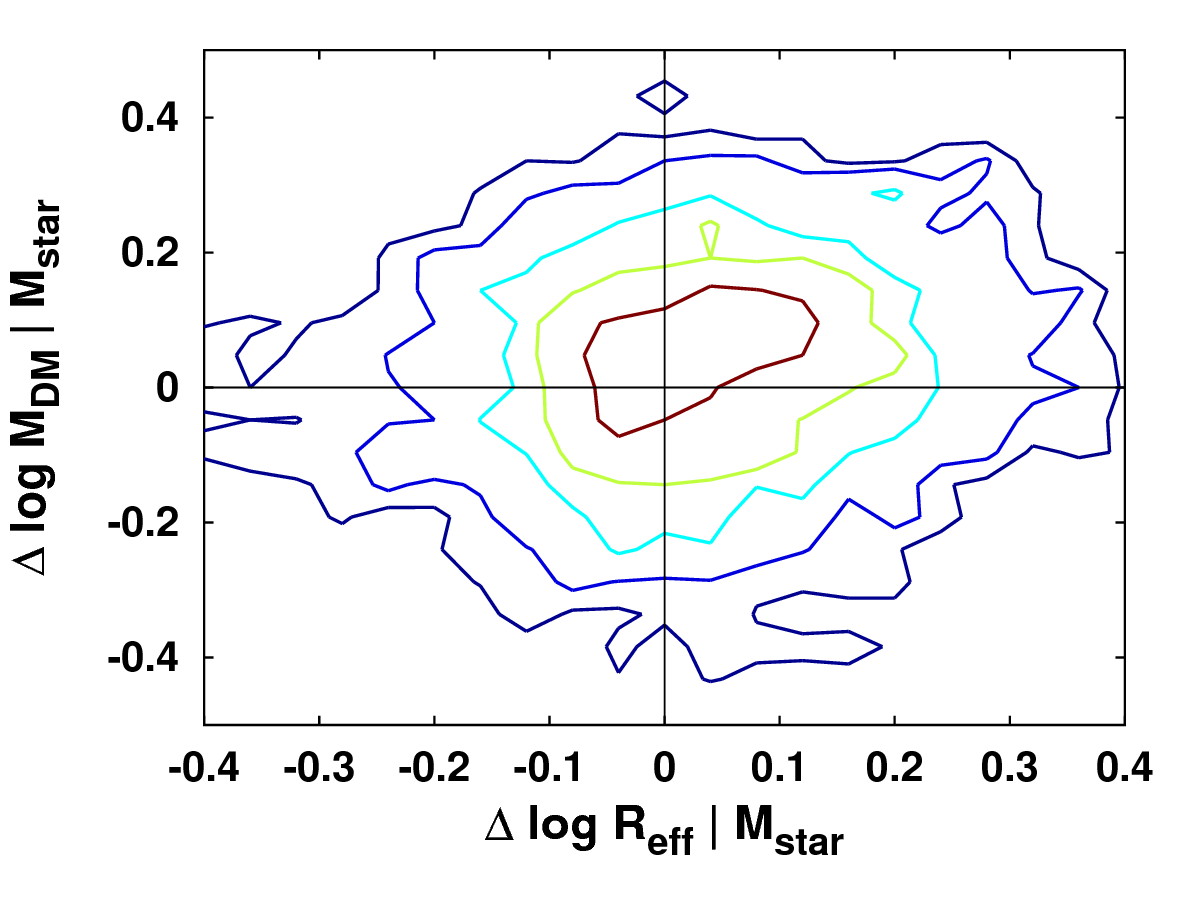}
    \label{fig:dRdM}
  }
  \subfigure[$\Delta R_\text{eff}-\Delta c$]
  {
    \includegraphics[width=0.33\textwidth]{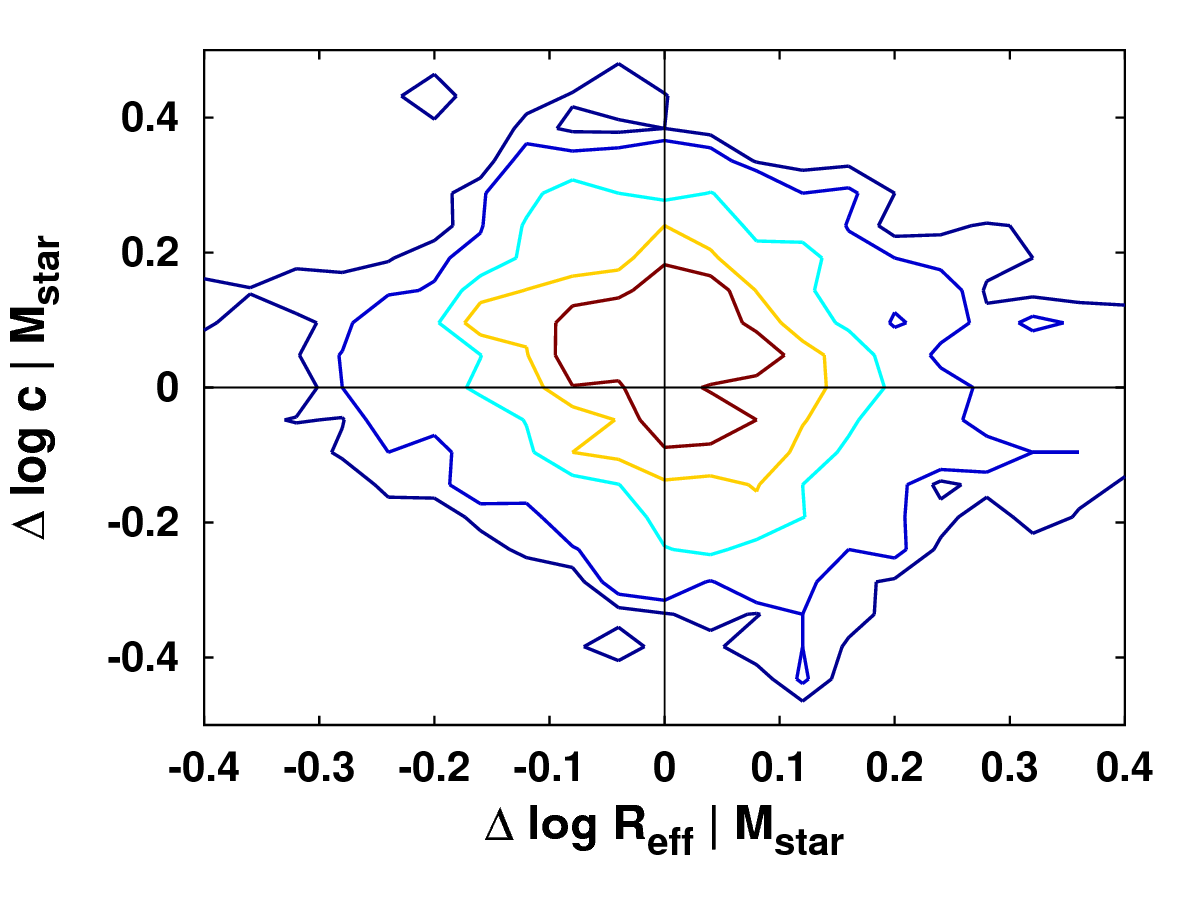}
    \label{fig:dRdc}
  }
  \subfigure[$\Delta R_\text{eff}-\Delta \lambda$]
  {
    \includegraphics[width=0.33\textwidth]{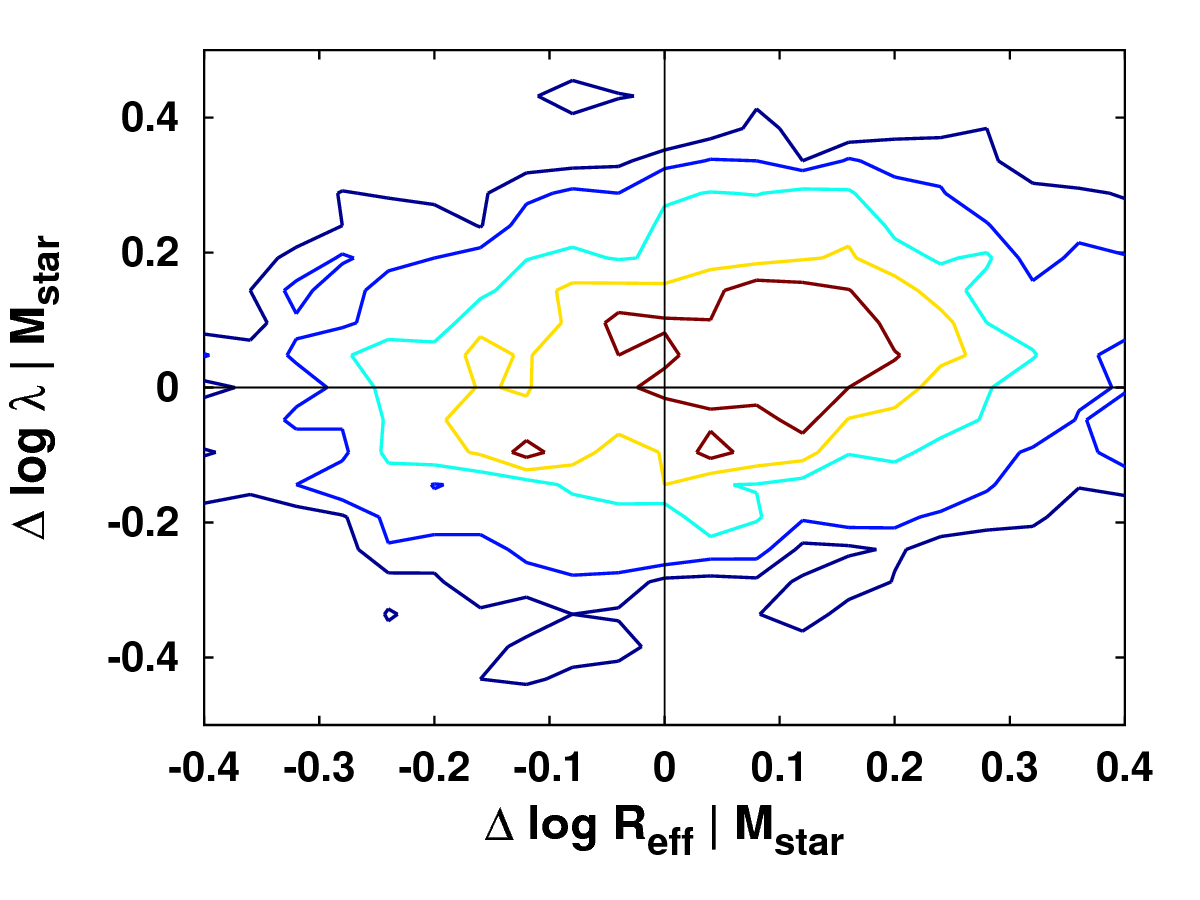}
    \label{fig:dRds}
  }
  \caption{The correlation of residuals of the stellar mass--size relation with dark matter mass, concentration and spin residuals in the hydro run of the \eagle{} simulation. Haloes were randomly selected from the catalogue to reproduce the stellar mass distribution of the P07 sample (see Section~\ref{sec:statistics}). In contrast to the SHAM+MMW model, \eagle{} predicts these correlations to be weak, which accounts for the better agreement of the predicted $s_\text{MSR}$ and $\rho_{\Delta R-\Delta V}$ values with the observations. The Spearman rank correlation coefficients of these relations are $0.18$, $-0.19$ and $0.17$, respectively (see Table~\ref{tab:table}).}
  \label{fig:dRdX}
\end{figure*}

Rows 3-5 of Table~\ref{tab:table} record the Spearman rank coefficients of the correlations between size residual ($\Delta R_\text{eff}$) and $M_\text{DM}$, $c$ and $\lambda$ residual in the \eagle{}, SHAM, and SHAM+MMW models. As halo properties cannot be observed, there are no corresponding entries in the first column. By construction, the SHAM model does not correlate galaxy size with any halo property at fixed stellar mass. As described in Section~\ref{sec:intro}, however, the SHAM+MMW model implies a strong correlation of $\Delta R_\text{eff}$ with $\Delta \lambda$ and a strong anticorrelation with $\Delta c$, and the latter in particular is responsible for the strongly negative value of $\rho_{\Delta R-\Delta V}$. In the \eagle{} simulation, galaxy size correlates only weakly with each halo variable, with the result that the predicted $\rho_{\Delta R-\Delta V}$ is similar to the SHAM case. In addition, this prevents $s_\text{MSR}$ from receiving the full contributions from the scatter in halo variables at fixed $M_*$, allowing it to remain below the P07 value. These correlations are shown explicitly in Figure~\ref{fig:dRdX}.

\section{Discussion and Conclusions}
\label{sec:conclusion}

While the principle component of the galaxy--halo connection -- the relation between galaxy mass and halo mass and concentration -- is becoming well constrained by abundance matching studies, secondary components, such as the dependence of galaxy size on halo properties, remain uncertain. A leading model for galaxy size (\citealt{MMW}; MMW) assumes $\sigma=0$ and proportionality of galaxy and halo specific angular momenta, making galaxy size a specific function of stellar mass and halo mass, concentration and spin. Despite success in matching the normalisation of the stellar mass--size relation when combined with SHAM, this model overpredicts both the spread in sizes ($s_\text{MSR}$) and the strength of the correlation of size and velocity residuals ($\rho_{\Delta R-\Delta V}$). This indicates a problem with the galaxy--halo connection it implies.

In this \textit{Letter}, we investigated this discrepancy in the context of the \eagle{} hydrodynamical simulation. We found the galaxy population in this simulation to exhibit near-agreement with measurements of both $s_\text{MSR}$ and $\rho_{\Delta R-\Delta V}$. We showed that this difference with the SHAM+MMW prediction is due not to modifications to the haloes themselves by baryons, but rather to the weakness of the correlations of galaxy size with $M_\text{DM}$, $c$ and $\lambda$. While the MMW model strongly correlates $R_\text{eff}$ with $M_\text{DM}$ ($\rho=0.75$), $c$ ($\rho=-0.76$) and $\lambda$ ($\rho=0.80$) at fixed $M_*$, the Spearman rank coefficients for the corresponding \eagle{} correlations are only $0.18$, $-0.19$ and $0.17$, respectively. These values are consistent with 0 within $3 \sigma$.

Our results have implications for both galaxy formation theory and semi-analytic and empirical modelling. On one hand, the breakdown of the MMW model requires explanation. Galaxy properties may become weakly correlated with halo spin due to stochastic transfer of angular momentum between baryons and dark matter, or a significant loss or redistribution through feedback or cooling processes~\citep{Brook,Illustris_AM}. On the other hand, the \eagle{} galaxy--halo correlations may be used to inform empirical models where galaxy sizes are added by hand. To match $s_\text{MSR}$ and $\rho_{\Delta R-\Delta V}$, at least in a SHAM framework, $R_\text{eff}$ should correlate at most weakly with $M_\text{DM}$, $c$ and $\lambda$ at fixed $M_*$. This is tacitly assumed by several existing models (e.g.~\citealt{D11, Dutton_13, DW16}), and implied also by aspects of the mass discrepancy--acceleration relation~\citep{D16}. We suggest such correlations be used by default from now on. Finally, our results facilitate the testing of galaxy formation theories: if a theory's effective galaxy--halo connection exhibits correlations compatible with those of \eagle{}, its success in matching the fine-grained statistics that we investigate here is assured.

\section*{Acknowledgements}

We thank Simon Foreman, Yu Lu, and Matthieu Schaller for comments on the manuscript, and Matthieu Schaller for guidance with the \eagle{} data. 
This work used the DiRAC Data Centric system at Durham University, operated by the Institute for Computational Cosmology on behalf of the STFC DiRAC HPC Facility (www.dirac.ac.uk). This equipment was funded by a BIS National E-infrastructure capital grant ST/K00042X/1, STFC capital grant ST/K00087X/1, DiRAC Operations grant ST/K003267/1 and Durham University. DiRAC is part of the National E-Infrastructure. HD received support from the U.S.\ Department of Energy, contract number DE-AC02-76SF00515. This work was supported by the Netherlands Organisation for Scientific Research (NWO), through VICI grant 639.043.409, and the European Research Council under the European Union's Seventh Framework Programme (FP7/2007-2013) / ERC Grant agreement 278594-GasAroundGalaxies. RAC is a Royal Society University Research Fellow.

\bsp

\end{document}